\documentclass[12pt]{article}

\usepackage{amsmath}
\usepackage{amsfonts}
\usepackage{amssymb}
\usepackage{graphics,graphicx,tikz}
\usepackage{soul}
\usepackage{tensor}
\usetikzlibrary{patterns}

\numberwithin{equation}{section} 
\usepackage[linktocpage]{hyperref}
\hypersetup{
	colorlinks=true,
	linkcolor=blue,
	filecolor=magenta,      
	urlcolor=blue,
	citecolor=blue
}
\def\beq{\begin{equation}}
\def\eeq{\end{equation}}
\newcommand{\commentOut}[1]{}
\def\bea{\begin{align}}
\def\eea{\end{align}}

\usepackage{color}

\begin{document}
	\hypersetup{pageanchor=false}
\begin{titlepage}
\hfill \hbox{NORDITA 2022-017}
\vskip 0.1cm
\hfill \hbox{UUITP-16/22}
\vskip 0.1cm
\hfill \hbox{QMUL-PH-22-12}
\vskip 1.5cm
\begin{flushright}
\end{flushright}
\vskip 1.0cm
\begin{center}
{\Large \bf Angular momentum of zero-frequency gravitons}
\vskip 1.0cm {\large  Paolo Di Vecchia$^{a, b}$, Carlo Heissenberg$^{b,c}$,
Rodolfo Russo$^{d}$} \\[0.7cm]

{\it \small $^a$ The Niels Bohr Institute, Blegdamsvej 17, DK-2100 Copenhagen, Denmark}\\
{\it \small $^b$ NORDITA, KTH Royal Institute of Technology and Stockholm University, \\
 Hannes Alfv{\'{e}}ns v{\"{a}}g 12, SE-11419 Stockholm, Sweden  }\\
 {\it \small $^c$ Department of Physics and Astronomy, Uppsala University,\\ Box 516, SE-75120 Uppsala, Sweden}\\
{\it \small $^d$ Centre for Theoretical Physics, Department of Physics and Astronomy,\\
Queen Mary University of London, Mile End Road, London, E1 4NS, United Kingdom.}
\end{center}

\begin{abstract}
By following closely Weinberg's soft theorem, which captures the $1/\omega$ pole contribution to the amplitude for soft graviton emissions ($\omega<\Lambda$) on top of an arbitrary background hard process, we calculate the expectation value of the graviton's angular momentum operator for arbitrary collisions dressed with soft radiation. We find that the result becomes independent of the cutoff $\Lambda$ on the graviton's frequency, effectively localizing at $\omega=0$. In this way, our result captures the contribution to the angular momentum that comes from the zero-frequency modes.  Like the soft theorem, our formula has an exact dependence on the kinematics of the hard particles and is only a function of their momenta.
As an example, we discuss in some detail the case of the $2 \to 2$ scattering of spinless particles in General Relativity and ${\cal N}=8$ supergravity.
\end{abstract}

\end{titlepage}
\hypersetup{pageanchor=true}

\tableofcontents

\section{Introduction}
\label{sec:intro}

The application of scattering-amplitude methods to the calculation of observables in classical gravity has led, in recent years, to the development of several new ideas and techniques. Owing to relativistic invariance, amplitudes have found a natural application to the problem of evaluating the Post-Minkowskian (PM) expansion of such observables. In this approach, which is well suited to the analysis of collision events at large impact parameter, gravitational interactions are taken to be weak while velocities are not assumed to be much smaller than the speed of light. Importing methods previously applied to integrand construction and to integral evaluation in the context of quantum amplitudes has proved pivotal in achieving progress in the PM analysis of the gravitational two-body dynamics  \cite{Goldberger:2004jt,Goldberger:2016iau,Damour:2017zjx,Luna:2017dtq,Cheung:2018wkq,Kosower:2018adc,Cristofoli:2019neg,Bjerrum-Bohr:2019kec,Mogull:2020sak,AccettulliHuber:2020dal,Jakobsen:2021smu,Cristofoli:2021vyo}. Recent results for collisions of spinless objects include the calculation of the 3PM and 4PM conservative deflection angle \cite{Bern:2019nnu,Bern:2019crd,Bern:2021dqo,Bern:2021yeh,Dlapa:2021vgp}, the full 3PM deflection angle \cite{Damour:2020tta,DiVecchia:2021ndb,DiVecchia:2021bdo,Bjerrum-Bohr:2021din,Brandhuber:2021eyq}, 3PM emitted energy and momentum \cite{Herrmann:2021lqe,Herrmann:2021tct,DiVecchia:2021bdo,Riva:2021vnj} and 3PM angular momentum \cite{Manohar:2022dea}. Several results obtained in this context have also been directly linked to the inspiral phase of merger events, via analytic continuation \cite{Kalin:2019rwq,Kalin:2019inp,Cho:2021arx}.

A prominent tool that has been helpful to extract classical information from the elastic $2\to2$ amplitude is the eikonal exponentiation \cite{Bjerrum-Bohr:2018xdl,KoemansCollado:2019ggb,Bern:2020gjj,Cristofoli:2020uzm,Parra-Martinez:2020dzs,AccettulliHuber:2020oou,DiVecchia:2020ymx,Bern:2020uwk,DiVecchia:2021ndb,Bern:2021dqo,DiVecchia:2021bdo,Bjerrum-Bohr:2021vuf},  whose inception actually dates back to the late eighties \cite{Amati:1987wq,tHooft:1987vrq,Amati:1987uf,Muzinich:1987in,Sundborg:1988tb,Amati:1990xe}.
Important endeavours have been also devoted to generalizing the eikonal framework to allow for the presence of additional outgoing graviton states \cite{Ciafaloni:2015xsr,Ciafaloni:2018uwe,Damgaard:2021ipf,Cristofoli:2021jas}, which are actually unavoidable in any physical process. 

However, as is well know from the literature on classical soft theorems \cite{Laddha:2018vbn,Sahoo:2018lxl,Saha:2019tub,Sahoo:2021ctw} (see \cite{Strominger:2017zoo} for a modern perspective on the connection between soft theorems and asymptotic charges), when focusing on the $\omega\to0$ limit in the spectrum of emitted radiation, general results become available and considerably simplify the analysis of the problem. Chief among them is Weinberg's soft graviton theorem \cite{Weinberg:1964ew,Weinberg:1965nx}, which relates the amplitude with a soft graviton emission to the one involving only hard states, up to a universal factor $F^{\mu\nu} = \kappa \sum_n p_n^\mu p_n^\nu / (p_n\cdot k)$ that only depends on the momenta of the hard states $p_n^\mu$ and on the graviton's momentum $k^\mu=\omega(1,\hat k)$. This formula is valid for any incoming and outgoing hard momenta, regardless of their specific properties, in particular their spin and mass---it even holds when some of the hard states are gravitons themselves. As such, the zero-frequency limit provides the ideal arena where one can hope to take a peek beyond the standard PM expansion, and investigate phenomena that take place even when its underpinning assumptions break down. Discussing the eikonal operator in the presence of soft graviton emissions and analysing the zero-frequency limit of the emitted energy spectrum, highlighting in particular its smoothness in the massless limit, is the main objective of the paper \cite{DiVecchia:2022nna}.

In this note, while leveraging on the framework proposed in \cite{DiVecchia:2022nna}, we instead focus on the contributions of soft gravitons to two quantities that have received a fair share of attention: the expression of the waveform in position space, in particular the value of the asymptotic shear in the far past, and the angular momentum  \cite{Damour:2020tta,Veneziano:2022zwh}. As we shall discuss, both quantities are sensitive to static field effects, and therefore to the inclusion of contributions due to ``gravitons with exactly zero frequency.'' In mathematical terms, this translates to the need of specifying an appropriate prescription on how to approach the $\omega=0$ singularity in $F^{\mu\nu}$. In the case of the waveform, it is indeed well known that $F^{\mu\nu}$ itself is proportional, after Fourier transform with respect to $\omega$, to the memory effect and to the asymptotic action of a BMS supertranslation \cite{Strominger:2014pwa}. This step relies on the fact that ``the Fourier transform of a pole in frequency space is a step function
in time.''~\cite{Strominger:2017zoo} In turn, it is precisely the supertranslation ambiguity that determines the value of the shear at early times and ultimately whether the leading contribution to the flux of angular momentum is of order $\mathcal O(G^2)$ or $\mathcal O(G^3)$, an issue recently discussed in \cite{Veneziano:2022zwh}.

Without the ambition of settling this delicate issue, in the following we shall follow a practical approach, mainly adopting Feynman's  prescription, which already plays a role in Weinberg's works \cite{Weinberg:1965nx,Weinberg:1995mt} (we elaborate on this point in Section~\ref{sec:eikopzfl}, for related discussions see \cite{Bautista:2021llr,Laddha:2018vbn,Bautista:2019tdr,Saha:2019tub,Manohar:2022dea}). As suggested by the soft eikonal operator \cite{DiVecchia:2022nna}, we also extend the soft theorem by applying it to the full \mbox{$S$-matrix} rather than to the connected \mbox{$T$-matrix} elements. Employing this simple albeit nonstandard recipe, we will for instance obtain a precise connection between $F^{\mu\nu}$ and the waveform \eqref{eq:wavAg} that selects its value at early retarded times.

In a similar fashion, after deriving a general formula linking the five-point amplitude to the angular momentum (also recently appeared in \cite{Manohar:2022dea}), applying it to $F^{\mu\nu}$ with the Feynman prescription will lead us to an explicit covariant expression for the angular momentum/mass dipole tensor $\mathcal J^{\alpha\beta}$ due to zero-frequency gravitons \eqref{Jgraviton}. Like the soft theorem, the validity of this formula is independent of the specific hard scattering process under consideration and only relies on the form of the Weinberg factor. When applied to $2\to2$ collisions and supplemented with the expression for the 1PM deflection angle, our formula reproduces exactly the $\mathcal O(G^2)$ results of \cite{Damour:2020tta,Manohar:2022dea}. We will provide below a detailed comparison with \cite{Jakobsen:2021smu,Mougiakakos:2021ckm,Gralla:2021qaf}, eventually finding complete agreement only with \cite{Manohar:2022dea} to this order. Our expression, once supplemented with the value for the 2PM deflection angle, also reproduces the part of the $\mathcal O(G^3)$ results of \cite{Manohar:2022dea} that is due to zero-frequency gravitons. However we emphasize once more that, just like Weinberg's theorem, the formula we provide holds independently of the specific background process, and in particular generalizes to any order in the PM expansion. 
In the same spirit, it holds for scattering and merger scenarios alike and also applies if the  hard states carry spin \cite{Alessio:2022kwv}.

Although the main target of current investigations remains general relativity, supergravity theories have also attracted interest \cite{Naculich:2008ew,Naculich:2011ry,White:2011yy,Akhoury:2011kq,Melville:2013qca,DiVecchia:2019myk,DiVecchia:2019kta,Bonocore:2020xuj,Bjerrum-Bohr:2021vuf,Bonocore:2021qxh}. In particular $\mathcal N=8$ supergravity has proved a useful theoretical laboratory for developing new tools and tackling conceptual challenges in a simpler setup \cite{Bern:2020gjj,Parra-Martinez:2020dzs,DiVecchia:2020ymx,DiVecchia:2021ndb,DiVecchia:2021bdo,Bjerrum-Bohr:2021vuf}.
For this reason, along the way, we provide the expression for Lorentz generators in terms of field oscillators in a form that is well-suited to amplitude applications not only for spin-2 (graviton) but also for spin-1 (vector) and spin-0 (scalar) fields (see also \cite{Campi1,Gonzo:2020xza,Manohar:2022dea}). Discussing these additional types of fields provides the ingredients that are needed to include  massless Kaluza--Klein scalars and vectors and the dilaton that eventually combine with the graviton to produce the full $\mathcal N=8$ result \eqref{JN8}.

A caveat is in order: because of the non-linear nature of gravitational interactions there are effects such as the non-linear memory \cite{Christodoulou:1991cr,Wiseman:1991ss,Blanchet:1992br} that are not directly captured by the Weinberg formula (see e.g.~the discussion in \cite{Damour:2020tta}). We leave the explicit analysis of this point from an amplitude perspective to future investigations, while remarking that instead such difficulties are absent for linear theories.

We start by quickly reviewing the method of dressed states and the soft eikonal operator \cite{DiVecchia:2022nna} in Sect.~\ref{sec:eikopzfl}, considering the derivation of the memory effect as a warm-up exercise for the introduction of the $-i0$ prescription. In Sect.~\ref{sec:angularmomentum} we discuss the expectation of the angular momentum operator in the state dressed by soft gravitons, obtaining our main formula Eq.~\eqref{Jgraviton}, which we then specialize to the case of $2\to2$ scattering in order to discuss its explicit features. 

We work with the mostly-plus metric $\eta_{\mu\nu} = \text{diag}(-1,+,+,\ldots,+)$ and with the convention
$A_{[\alpha}B_{\beta]}\equiv A_{\alpha}B_{\beta}-A_{\beta}B_{\alpha}$ without symmetry factors. 

\section{The Eikonal Operator in the ZFL}
\label{sec:eikopzfl}

Our framework is the eikonal operator for soft  radiation introduced in \cite{DiVecchia:2022nna}, where the methods of  Block-Nordsieck~\cite{Bloch:1937pw,Thirring:1951cz} and the closely related approach by Weinberg~\cite{Weinberg:1964ew,Weinberg:1965nx} are adapted to the discussion of the classical scattering problem (see also \cite{Mirbabayi:2016axw,Choi:2017ylo,Arkani-Hamed:2020gyp}). The emission of these soft quanta exponentiates in momentum space, so we write the $S$-matrix dressed with soft-radiation  $S^{(M)}_{s.r.}$ in terms of that of the hard process $S^{(M)}$
\begin{equation}
  \label{eq:eiksr}
  \begin{split}
   S_{s.r.}^{(M)} = \frac{e^{2i\hat\delta_{s.r.}} S^{(M)}}{\langle0|e^{2i\hat \delta_{s.r.}}|0\rangle}\,,
   \qquad
   2i\hat\delta_{s.r.}
   =
   \frac1\hbar\int_{\vec k} \sum_{j} \left[f_j(k) a^\dagger_j(k)- f^\ast_j(k) a_j(k)\right],
  \end{split}
\end{equation}
where $j$ runs over the physical polarisations, whose polarisation ``tensors'' are $\varepsilon_j^{\mu\nu}(k)$, and we work with the conventions
\begin{equation}\label{}
	[ a^{\phantom{\dagger}}_i (k) , a_j^\dagger (k')] =\delta(\vec k,\vec k\,')\delta_{ij}
\end{equation}
and
\begin{equation}
	\label{eq:aadcom}
	\delta(\vec k,\vec k') =  2 \hbar \omega (2\pi)^{D-1}  
	\delta^{(D-1)} 
	(\vec{k}-\vec{k}') 
	\,,
	\qquad
	\int_{\vec k} \equiv \int \frac{d^{D-1}\vec k}{2 \omega(2\pi)^{D-1}}\,,
\end{equation}
where $k^0=|\vec k|=\omega$. For instance, in a classical $2\to 2$ scattering process, one can derive the $S^{(M)}$ to be inserted in \eqref{eq:eiksr} by taking the Fourier transform of the elastic eikonal $e^{i2\delta_0}$ from the impact parameter, $b$, to momentum space
\begin{equation}
  \label{eq:btoq}
  S^{(M)}(\sigma,Q) \simeq \int d^{D-2} b\,  e^{-i bQ} e^{2i\delta_0(\sigma,b)}\;,
\end{equation}
which can be evaluated as usual by a saddle-point approximation. 

The quantity $f_j(k)$ in~\eqref{eq:eiksr} is determined by the Weinberg soft factor for the emission of a soft graviton on top of the background hard process. However, this is not a standard application of Weinberg's theorem, which only holds for {\em connected} amplitudes \cite{Weinberg:1995mt,Weinberg:1995mt}, while here we are dressing the full $S$-matrix, including the identity term. 
Moreover, for our discussion, it will be important to retain the $-i0$ prescription that is dictated by the Feynman propagator to which the soft particle is attached, so that
\begin{equation}\label{eq:SMNSM}
	f_j(k) = \varepsilon^{\ast\mu\nu}_j(k)F_{\mu\nu}(k)\,,\qquad
	F^{\mu\nu}(k) = \sum_{n}\frac{\kappa\, p_n^\mu p_n^\nu}{p_n\cdot k-i0}\,.
\end{equation}
Note that we work with conventions according to which the vector $k^\mu$ is always future directed, $k^0>0$, while all momenta $p_n$ are regarded as outgoing so that 
\begin{equation}\label{}
	p_n =\eta_n (E_n, \vec k_n)\,,\qquad E_n>0\,,
\end{equation}
where $\eta_n$ takes the value $+1$ for outgoing and $-1$ for incoming states.
In Weinberg's approach the Feynman prescription is included in the general form of the soft factor (Sect.~II.1 and II.2 of \cite{Weinberg:1965nx}, Sect.~13.1 of \cite{Weinberg:1995mt}), but this only plays a direct role in the contributions of virtual soft particles to infrared divergences (Sect.~II.3 and V of \cite{Weinberg:1965nx}, Sect.~13.2 of \cite{Weinberg:1995mt}), while being irrelevant for the calculation of real soft particle and energy emission rates (Sect.~II.4 and V of \cite{Weinberg:1965nx}, Sect.~13.3 of \cite{Weinberg:1995mt}). In order to study the gravitational field, including static effects that are relevant to the calculation of angular momentum, we shall see below that the $-i0$ is also important in our setup to resolve the singularity lying at the $\omega=0$ end of the spectrum of real gravitons.

The choice of $-i0$ prescription in \eqref{eq:SMNSM} can be further supported by considering the stress-energy tensor $\mathcal T^{\mu\nu}(x)$ for an idealized scattering process localized at the origin of space-time
\cite{Weinberg:1972kfs,Garfinkle:2017fre,Campoleoni:2019ptc}, 
\begin{equation}\label{stressenergy}
	\mathcal T^{\mu\nu}(x) = \sum_n  \theta(\eta_n x^0)\,  m^{\phantom{\mu}}_n u_n^{\mu}u_n^{\nu}\int_{-\infty}^{+\infty}\delta^{(D)}(x-u_n \tau)\,d\tau\,,
\end{equation}
where
\begin{equation}\label{}
	p_n^\mu = \eta_n m_n u^\mu_n\,,\qquad u^\mu_n\, u_{n,\mu}=-1\,,\qquad u^0_n>0\,.
\end{equation}
$\mathcal T^{\mu\nu}(x)$ is conserved $\partial_\mu \mathcal T^{\mu\nu}(x)=0$ thanks to 
\begin{equation}\label{}
	\sum_{n} p_n^\mu = 0\,,
\end{equation}
and going to Fourier space on gets precisely
\begin{equation}\label{}
	\tilde{\mathcal{T}}^{\mu\nu}(k) = \int d^Dx\,\mathcal T^{\mu\nu}(x)\,e^{-ik\cdot x} 
	= \frac{1}{i\kappa}\, F^{\mu\nu}(k)
\end{equation}
with the $-i0$ prescription as in \eqref{eq:SMNSM}. 

A crucial point for us is that $F^{\mu\nu}$ retains an exact dependence on the momenta of the background hard process, no matter whether deflections  are large or small, and is insensitive to the detailed structure of the hard states themselves.
In the following, we will apply the eikonal operator to discuss the contribution of low-energy gravitons to two observables: the waveforms and the angular momentum. The general strategy, given any quantum observable $\mathcal O$, is to take its expectation value according to
\begin{equation}\label{SOS}
	\langle \mathcal O \rangle = \langle 0| S_{s.r.}^\dagger \mathcal O \, S_{s.r.} |0\rangle\,, 
\end{equation}
where $S_{s.r.}$ is the eikonal operator in $b$-space \cite{DiVecchia:2022nna}.
Physically, this represents the average value of $\mathcal O$ in the final state of the scattering event, obtained by applying $S_{s.r.}$ to the state with no gravitons.

We conclude this section by mentioning that the soft eikonal operator is easily modified to accommodate for the presence of other massless fields (scalars and vectors), which intervene for instance in $\mathcal{N}=8$ supergravity where the massive particles are described by Kaluza--Klein (KK) modes. More precisely, we consider 10D massless scalars that acquire masses $m_1$ and $m_2$ via KK compactification in two orthogonal directions (see \cite{Caron-Huot:2018ape}; here we work with the $\sin\phi=1$ case of~\cite{Parra-Martinez:2020dzs}; see also~\cite{DiVecchia:2021ndb}). Then we have
\begin{equation}\label{fvecfsc}
	f_j^\text{vec}(k)=\sum_{n} \eta_n e_n \frac{\varepsilon^\ast_{\mu,j}(k)\, p_n^\mu}{p_n\cdot k-i0},\qquad
	f^\text{sc}(k) = \sum_{n} \frac{g_n}{p_n\cdot k-i0}
\end{equation}
for vectors and scalars respectively, where $e_n$ and $g_n$ denote suitable couplings.
Specifically, for the dilaton, $g_n = -\kappa\, m_n^2/\sqrt{D-2}$, for KK vectors $e_n=\sqrt 2\,\kappa\, m_i$ and for KK scalars $g_n =\kappa\, m_i^2$. The exponentiation of soft quanta then works as in \eqref{eq:eiksr}, replacing $2i\hat\delta_{s.r.}$ by
\begin{align}
    2i\hat{\delta}^{\mathcal{N}=8}_{s.r.} & =\frac{1}{\hbar}\int_{\vec k} \sum_{j} 
    \left[ (f_j a^\dagger_j-f^\ast_j a_j) + (f^{d} a_d^\dagger- f^{d\ast}a_d) +  \right. \nonumber \\ & ~ + \left. (f^{v}_{j} a^\dagger_{v,j}-f^{v\ast}_{j} a_{v,j}) + (f^{s}_{j} a^\dagger_{s,j}-f^{s\ast}_{j} a_{s,j}) \right].   \label{eq:dn8sr}   
\end{align}

\subsection{Memory effect}
\label{ssec:wavzfl}

As a warm up, let us start by checking how our formulas reproduce the well-known connection between the Weinberg soft theorem and the memory effect \cite{Strominger:2014pwa,Strominger:2017zoo}.
The classical field is obtained by inserting in the expectation value \eqref{SOS} the Fock field operator
\begin{equation}\label{FockH}
	H_{\mu\nu}(x)=\int_{\vec k} \left[ a_{\mu\nu}(k)\, e^{ikx} + a_{\mu\nu}^\dagger(k)\,e^{-ikx}\right],
	\qquad
	\langle H_{\mu\nu}(x) \rangle = h_{\mu\nu}(x)\,,
\end{equation}
where for definiteness we work with the field in De Donder gauge,\footnote{We follow the standard amplitude nomenclature and call De Donder gauge the gravity analogue of Feynman gauge in electromagnetism. In this approach, the components of $H_{\mu\nu}$ are regarded as independent. This should not be confused with the classical terminology according to which De Donder gauge amounts to imposing $2\partial_\nu h^{\mu\nu} = \eta^{\alpha\beta}\partial^\mu h_{\alpha\beta}$ on the metric fluctuation.} 
\begin{equation}\label{DDCCR}
	[a_{\mu\nu}(k), a^\dagger_{\rho\sigma}(k')] =\delta(\vec k,\vec k\,')\, \frac12\left[
	\eta_{\mu\rho}\eta_{\nu\sigma}
	+
	\eta_{\mu\sigma}\eta_{\nu\rho}
	-
	\frac{2}{D-2}\,\eta_{\mu\nu}\eta_{\rho\sigma}
	\right].
\end{equation}
This yields,
\begin{equation}
	\label{eq:hmnzfl}
	h_{\mu\nu}(x)=\int_{\vec k}  \,\langle 0| S_{s.r.}^\dagger  \left[ a_{\mu\nu}(k)\, e^{ikx} + a_{\mu\nu}^\dagger(k)\,e^{-ikx}\right] S_{s.r.}|0\rangle\,.
\end{equation}
Using
\begin{equation}\label{fTT}
	f^{\mu\nu}(k) = \Pi^{\mu\nu}_{\rho\sigma}(\hat k) F^{\rho\sigma}(k)\,,
	\qquad
	F^{\mu\nu}(k)
	=
	\sum_n \frac{\kappa\,  p_n^\mu p_n^\nu}{p_n\cdot k-i0}\,,
\end{equation}
where $\Pi^{\mu\nu}_{\rho\sigma}$ is the usual transverse-traceless projector over physical degrees of freedom (see \eqref{TTprojector}), together with
\begin{equation}\label{key}
	\sum_{j} f_j(k) a_j^\dagger(k)  = f^{\mu\nu}(k) a^\dagger_{\mu\nu}(k)\,,
\end{equation}
and similarly for $a_j$,
this leads to
\begin{equation}
	\label{eq:hmnzfz}
	h_{\mu\nu}(x) = \int_{\vec k} \left[    f_{\mu\nu}(k)  \,e^{ikx} +   f^\ast_{\mu\nu}(k) \,e^{-ikx}\right].
\end{equation}

Now we consider the asymptotic limit for the gravitational field, where $x^\mu = (x^0,\vec x) = (u+r,r\hat x)$ and the detector's distance is taken to infinity, $r\to\infty$, for fixed retarded time $u$ and angles $\hat x$.
In this limit, a standard stationary-phase argument (see e.g.~\cite{Strominger:2017zoo}) yields
\begin{equation}\label{}
	h_{\mu\nu} (u+r,r \hat x) \sim \frac{1}{4i\pi r}
	\int_0^\infty \frac{d\omega}{2\pi}\, 
	\left[
	f_{\mu\nu}(\omega,\omega \hat x)\, e^{-i\omega u} 
	-
	f_{\mu\nu}^\ast(\omega,\omega \hat x)\, e^{i\omega u} 
	\right],
\end{equation}
where  we have focused on the $D=4$ case.
Note that in this step the angular integral over the momenta $\hat k$ localizes along the observation direction $\hat x$.
Recalling $f_{\mu\nu}(k)=-f_{\mu\nu}^\ast(-k)$ then leads to
\begin{equation}\label{hmunufmunu}
	h_{\mu\nu} \sim 
	\frac{1}{4i\pi r}
	\int_{-\infty}^{+\infty} \frac{d\omega}{2\pi}\, 
	f_{\mu\nu}(\omega,\omega \hat x)\, e^{-i\omega u} \,,
\end{equation}
where the two terms recombined to reconstruct a single integral over positive and negative frequencies \cite{Cristofoli:2021vyo}.
Letting $p_n = \eta_n(E_n,\vec k_n)$ with $E_n>0$, we are thus left with integrals of the type
\begin{equation}\label{}
	\int_{-\infty}^{+\infty} \frac{d\omega}{i2\pi} \frac{e^{-i\omega u}}{-\eta_n \omega-i0}= \int_{-\infty}^{+\infty} \frac{d\omega}{i2\pi}\frac{e^{i\omega \eta_n u}}{ \omega-i0} = \theta(\eta_n u)\,,
\end{equation}
with the $\theta$ the Heaviside step function.
In this way, we have
\begin{equation}\label{hTT}
	h^{\mu\nu} \sim \frac{\kappa}{4\pi r} \Pi^{\mu\nu}_{\rho\sigma}(\hat x)\,\sum_{n} \frac{p_n^\rho p_n^\sigma\, \theta(\eta_n u)}{E_n-\vec k_n\cdot\hat x}\,.
\end{equation}
Adjusting the overall normalisation by comparing 
\begin{equation}\label{}
	g_{\mu\nu} = \eta_{\mu\nu} + 2 W_{\mu\nu}  = \eta_{\mu\nu} + 2\kappa h_{\mu\nu}\,,
\end{equation}
we find the waveform
\begin{equation}
	\label{eq:wavAg}
	W^{\mu\nu} = \kappa h^{\mu\nu} \sim  \frac{2G}{r} \Pi^{\mu\nu}_{\rho\sigma}(\hat x) \sum_{n} \frac{p_n^\rho p_n^\sigma\, \theta(\eta_n u)}{E_n-\vec k_n\cdot\hat x}\,,
\end{equation}
or more explicitly
\begin{equation}\label{}
	W^{\mu\nu} \sim  \frac{2G}{r} \Pi^{\mu\nu}_{\rho\sigma}(\hat x) \left[
	\theta(u)\sum_{\text{out}} \frac{p_n^\rho p_n^\sigma}{E_n-\vec k_n\cdot\hat x}
	+
	\theta(-u)\sum_{\text{in}} \frac{p_n^\rho p_n^\sigma}{E_n-\vec k_n\cdot\hat x}
	\right].
\end{equation}
For $u<0$ only the second term in the square brackets survives and determines the value of the asymptotic shear in the far past. In fact, expanding this expression in the PM regime, where the out states are equal to the in states up to $\mathcal O(G)$ deflections, one finds 
\begin{equation}\label{Damourf1}
		W^{\mu\nu} \sim  \frac{2G}{r} \Pi^{\mu\nu}_{\rho\sigma}(\hat x) \sum_{\text{in}} \frac{p_n^\rho p_n^\sigma}{E_n-\vec k_n\cdot\hat x}
	+\mathcal O(G^2)\,,
\end{equation}
i.e.~the waveform has a $u$-independent $\mathcal O(G)$ contribution. Clearly this term arises by the PM expansion of the soft factor acting on the identity in \eqref{eq:eiksr}, hence the need of considering a soft dressing of the full $S$-matrix in our approach. As a result, Eq.~\eqref{Damourf1} agrees in particular with \cite{Damour:2020tta}, where the term \eqref{Damourf1} is eventually instrumental in the derivation of an $\mathcal O(G^2)$ flux of angular momentum.
Eq. \eqref{Damourf1} is the static gravitational field generated  by a set of free particles that do not undergo any deflection to leading order in $\mathcal O(G)$.

Let us also mention that modifying the $-i0$ prescription according to 
\begin{equation}\label{modi0}
	F^{\mu\nu} = \sum_{n}\frac{\kappa \, p_n^\mu p_n^\nu}{p_n\cdot k-\eta_n i0}\,,
\end{equation}
instead of the Feynman prescription adopted in Eq.~\eqref{fTT}, one instead obtains
\begin{equation}\label{Wmod}
	W^{\mu\nu} \sim \frac{2G}{r}\,\theta(u) \Pi^{\mu\nu}_{\rho\sigma}(\hat x)\,\sum_{n} \frac{p_n^\rho p_n^\sigma \, \eta_n}{E_n-\vec k_n\cdot\hat x}\,.
\end{equation}
The difference between \eqref{eq:wavAg} and \eqref{Wmod} is precisely the ($u$-independent) ambiguity under BMS supertranslations, which was recently discussed in detail in \cite{Veneziano:2022zwh} in connection with the definition of the angular momentum. According to the terminology introduced in that reference, Eq.~\eqref{Wmod} is the waveform in the canonical Bondi frame, while Eq.~\eqref{eq:wavAg} is the waveform in the intrinsic one (i.e.~the one employed in \cite{Bini:2012ji,Damour:2020tta,Bini:2021gat}).

Both \eqref{eq:wavAg} and \eqref{Wmod} reproduce the leading result of~\cite{Laddha:2018vbn,Sahoo:2018lxl,Saha:2019tub,Sahoo:2021ctw} by considering the supertranslation-invariant combination 
\begin{equation}\label{}
	W_{\mu\nu}(u>0)-W_{\mu\nu}(u<0)
	=\frac{2G}{r}\,\Pi^{\mu\nu}_{\rho\sigma}(\hat x)\,\sum_{n} \frac{p_n^\rho p_n^\sigma \, \eta_n}{E_n-\vec k_n\cdot\hat x}\,.
\end{equation}
This is the term indicated as $A_{\mu\nu}$ in~\cite{Sahoo:2021ctw}, and can be itself regarded as the action of a supertranslation \cite{Strominger:2014pwa}.

We conclude this section by quoting the results for the corresponding waveforms associated to vector and scalar radiation, which are relevant in particular for the $\mathcal N=8$ setup. The field operators can be conveniently taken as
\begin{align}\label{FockA}	
	A_\mu(x) 
	&= 
	\int_{\vec k} \left[
	a_\mu(x) \, e^{ik\cdot x} + a_\mu^\dagger(x) \, e^{-ik\cdot x}
	\right],
	\\
	\label{FockPhi}
	\Phi(x)
	&=
	\int_{\vec k} \left[
	a(x) \, e^{ik\cdot x} + a^\dagger(x) \, e^{-ik\cdot x}
	\right]
\end{align}
where 
\begin{equation}\label{}
	[a_{\mu}(k), a^\dagger_{\nu}(k')] = \delta(\vec k,\vec k\,')\, \eta_{\mu\nu}\,,
	\qquad
	[a(k), a^\dagger(k')] = \delta(\vec k,\vec k\,')\,,
\end{equation}
choosing Feynman gauge for the vector. Taking expectation values
as in \eqref{SOS}, and evaluating the asymptotic limit as $r\to\infty$ for fixed $u,\hat x$, leads to
\begin{equation}\label{}
	\langle A_\mu(x) \rangle \sim 
	\frac{1}{4\pi r} \Pi^{\mu}_{\nu}(\hat x)\,\sum_{n} \frac{e_n\,p_n^\nu\, \theta(\eta_n u)}{E_n-\vec k_n\cdot\hat x}\,,
	\quad
	\langle \Phi(x) \rangle \sim 
	\frac{1}{4\pi r} \sum_{n} \frac{g_n\, \theta(\eta_n u)}{E_n-\vec k_n\cdot\hat x}
\end{equation}
where $\Pi^{\mu\nu}$ is the transverse projector (see \eqref{Pimunu}).

\section{Angular Momentum}
\label{sec:angularmomentum}

While the angular momentum flux can be directly evaluated starting from the asymptotic waveform, \cite{Peters:1964zz,DeWitt:2011nnj,Thorne:1980ru,Bonga:2018gzr,Blanchet:2018yqa} (see e.g. \cite{Maggiore:2007ulw} for a textbook presentation), in this approach care must be exerted concerning the inclusion of the so-called static or Coulombic mode and to ambiguities related to a redefinition of the ``angular momentum'' aspect \cite{Ashtekar:2017ydh,Ashtekar:2017wgq,Bonga:2018gzr,Compere:2019gft}. In the following, we follow an approach more closely related to scattering-amplitude calculations \cite{Kosower:2018adc,Cristofoli:2021vyo,Manohar:2022dea}.
In particular, continuing the approach of \cite{DiVecchia:2022nna}, we simply insert in \eqref{SOS} the angular momentum generator and derive the resulting expressions for scalars, gravitons and vectors. 

\subsection{Scalar}

To simplify matters, let us start by focusing on the case of the massless scalar field. In this case, one can obtain the relevant operator by inserting the Fock expansion \eqref{FockPhi} in the charge associated to Lorentz generators derived from the standard Noether method \cite{Peskin:1995ev,Campi1}. This leads to the following expression
\begin{equation}\label{Jsc}
	J_{\alpha\beta}^\text{sc} =-\frac{i}{2}\int_{\vec k} \left[
	a^\dagger(k) 
	k^{\phantom{[}}_{[\alpha} \frac{\partial 	a(k)}{\partial k^{\beta]}}
	-
	k^{\phantom{[}}_{[\alpha} 
	\frac{\partial 	a^\dagger(k)}{\partial k^{\beta]}}
	\,
	a(k) 
	\right]
	\equiv
	-i\int_{\vec k} 
	a^\dagger(k) 
	k^{\phantom{[}}_{[\alpha} \frac{\overset{\leftrightarrow}{\partial} a(k)}{\partial k^{\beta]}}
\,.
\end{equation}
Note that we include a factor of $\tfrac12$ in the notation $\overset{\leftrightarrow}{\partial}$, while we work with the convention
 $A_{[\alpha}B_{\beta]}\equiv A_{\alpha}B_{\beta}-A_{\beta}B_{\alpha}$ without symmetry factors. 
 
In Eq.~\eqref{Jsc}, although the measure is on shell, the ladder operators $a(k)$ are regarded as functions with argument $k^\mu$. If we instead worked with the ladder operators $a(\vec k)$ that are functions of $\vec k$ only, as perhaps more customary in textbook discussions \cite{Peskin:1995ev,Campi1}, we would have $a(|\vec k|,\vec k)= a(k)$ with $k^0=|\vec k|$ and letting $I,J=1,2,\ldots,D-1$ label the spatial components,
\begin{equation}
	\frac{\partial a(\vec k)}{\partial k^I} = \frac{k^I}{k^0} \frac{\partial a(k)}{\partial k^0} + \frac{\partial a(k)}{\partial k^I}\,.
\end{equation}
In this way, Eq.~\eqref{Jsc} would translate to
\begin{align}\label{J}
	J_{IJ}^\text{sc} = - i \int_{\vec k} a^\dagger(\vec k) 
	k^{\phantom{[}}_{[I} \frac{\overset{\leftrightarrow}{\partial}}{\partial k^{J]}}\,
	a(\vec k)\,,\qquad
	J_{0I}^\text{sc} =  - i \int_{\vec k} k^{\phantom{[}}_{0}\, a^\dagger(\vec k) \,
	 \frac{\overset{\leftrightarrow}{\partial}}{\partial k^I}\,
	a(\vec k)\,,
\end{align}
where in the last expression $k_0=-k^0=-|\vec k|$. Note the absence of explicit antisymmetrisation with respect to $0I$ in this latter way of writing $J_{0I}^{\text{sc}}$.

We then insert \eqref{Jsc} in \eqref{SOS} and define the average
\begin{equation}\label{}
	\mathcal J^\text{sc}_{\alpha\beta} = \langle J^\text{sc}_{\alpha\beta}\rangle\,.
\end{equation}
Like in the previous section, we shall follow the Feynman prescription to approach the $\omega=0$ singularity:
\begin{equation}\label{}
	\mathcal J^\text{sc}_{\alpha\beta} 
	=
	-
	\frac{i}2
	\int_{\vec k} \left(f^\ast k_{[\alpha} \frac{\partial f}{\partial k^{\beta]}} - k_{[\alpha} \frac{\partial f^\ast}{\partial k^{\beta]}} f \right),\qquad
	f = \sum_{n}\frac{g_n}{p_n\cdot k-i0}\,.
\end{equation}
Substituting, and making the upper cutoff $\Lambda$ explicit, we get
\begin{equation}\label{}
	(\mathcal J^\text{sc})^{\alpha\beta} 
	=
	\frac{i}{2} \sum_{n,m}g_n g_m \!\int_{\vec{k}}
	\left(\frac{k_{\phantom{m}}^{[\alpha}p_m^{\beta]}\,\theta(\Lambda-k^0)}{(p_n\cdot k+i0)(p_m\cdot k-i0)^2} - \frac{k_{\phantom{n}}^{[\alpha}p_n^{\beta]}\,\theta(\Lambda-k^0)}{(p_n\cdot k+i0)^2(p_m\cdot k-i0)} \right)\!.
\end{equation}
It is now convenient to change variable according to $k = \ell$ in the first term and $k=-\ell$ in the second term, as well as relabel $m\leftrightarrow n$ in the second term.
Introducing $\operatorname{sgn}(\ell^0)=\theta(\ell^0)-\theta(-\ell^0)$, this can be then written more compactly as
\begin{equation}\label{vartheta}
	(\mathcal J^\text{sc})^{\mu\nu}
	=
	\frac{i}{2} \sum_{n,m}g_n g_m \int\frac{d^4\ell}{(2\pi)^4}  
	\frac{\operatorname{sgn}(\ell^0)2\pi\delta(\ell^2)\ell_{\phantom{m}}^{[\mu}p_m^{\nu]}\,\theta(\Lambda-|\ell^0|)}{(p_n\cdot \ell+i0)(p_m\cdot \ell-i0)^2}\,,
\end{equation}
where we restricted to $D=4$.

We need to calculate the following integral,
\begin{equation}\label{Kmu}
	K^\mu = \int\frac{d^4\ell}{\pi^2}  
\frac{\operatorname{sgn}(\ell^0)2\pi\delta(\ell^2)\theta(\Lambda-|\ell^0|)\ell^{\mu}}{(p_n\cdot \ell+i0)(p_m\cdot \ell-i0)^2} = A p_n^\mu + B p_m^\mu\,,
\end{equation}
where we have used the fact that $K^\mu$ transforms as a Lorentz vector	in spite of the cutoff. Indeed, as we shall see shortly, the cutoff dependence drops out thanks to the scale invariance of the integrand and the symmetric cutoff imposed by the $\theta$ function (see also \cite{DiVecchia:2022nna} for a more detailed discussion). 
As a matter of fact, only the contribution proportional to $p_n^\mu$ enters the calculation, owing to antisymmetrisation  with $p_m^{\nu}$ in \eqref{vartheta}, so we don't need the coefficient $B$.
In turn, $A$ can be obtained from
\begin{equation}\label{A}
A = \frac{m_m^2 p_n\cdot K + (p_n\cdot p_m)p_m\cdot K}{(p_n\cdot p_m)^2-m_n^2m_m^2}\,.
\end{equation}

The remaining scalar integrals are almost identical to the ones discussed e.g. in the Appendix B of \cite{Heissenberg:2021tzo}. After using the delta function in the measure, one is left with an integral over the graviton's direction $\hat k$ and over $\omega=\pm|\vec k|$. The angular integration can be performed using \cite{Weinberg:1965nx,Heissenberg:2021tzo}
\begin{equation}\label{}
	\oint \frac{d\Omega_2(\hat k)}{(E_n-\vec k_n \cdot \hat k)(E_m-\vec k_m \cdot \hat k)} = \frac{4\pi \Delta_{nm}}{m_n m_m}\,,
\end{equation}
where we recall that $p_n=\eta_n(E_n, \vec k_n)$, with $\eta_n=+1$ if $n$ is outgoing and $\eta_n=-1$ if $n$ is incoming, and we introduced the shorthand notation
\begin{equation}\label{sigmasDeltas}
	\sigma_{nm}=-\eta_n\eta_m \, \frac{p_n\cdot p_m}{m_n m_m}\,,
	\qquad
	\Delta_{nm} = \frac{\operatorname{arccosh}\sigma_{nm}}{\sqrt{\sigma^2_{nm}-1}}\,.
\end{equation}
Moreover, the integral over $\omega$ can be evaluated using\footnote{
One can directly evaluate the integrals $\int_{-\Lambda}^{\Lambda} \frac{\omega\,d\omega}{(-\eta_n\omega+i\lambda)(-\eta_m\omega-i\lambda)}$ and 
$\int_{-\Lambda}^{\Lambda} \frac{\omega\,d\omega}{(-\eta_m\omega-i\lambda)^2}$ for fixed $\Lambda>\lambda>0$, and then send $\lambda\to0^+$ to obtain the desired results.
}
\begin{align}\label{omegaintnm}
	\int_{-\Lambda}^{\Lambda} \frac{\omega\,d\omega}{(-\eta_n\omega+i0)(-\eta_m\omega-i0)} &=  -\frac{i\pi}{2}(\eta_n-\eta_m)\,,\\
	\label{omegaintmm}
	\int_{-\Lambda}^{\Lambda} \frac{\omega\,d\omega}{(-\eta_m\omega-i0)^2} &= -i\pi\eta_m\,.
\end{align}
Note that in this step the cutoff dependence drops out, indicating that the integration effectively localizes at $\omega=0$.
As a result
\begin{equation}\label{pnKint}
	p_n\cdot K
	=
	-\frac{4i\pi\eta_m}{m_m^2}
	\,,\qquad
	p_m\cdot K = \frac{2i\pi\eta_n\eta_m\Delta_{nm}}{m_nm_m} (\eta_n-\eta_m)\,.
\end{equation}
Substituting into \eqref{A},
\begin{equation}\label{}
	A = -2i\pi\frac{\sigma_{nm}\Delta_{nm}(\eta_{n}-\eta_m)+2\eta_m}{m_n^2m_m^2(\sigma_{nm}^2-1)}\,,
\end{equation}
and thus \eqref{Kmu} and \eqref{vartheta} combine to yield
\begin{equation}\label{Jradscalar}
	(\mathcal J^{\text{sc}})^{\mu\nu} = \frac{1}{16\pi} \sum_{n,m}\frac{g_ng_m}{m_n^2m_m^2} \frac{\sigma_{nm}\Delta_{nm}-1}{\sigma_{nm}^2-1} (\eta_n-\eta_m)p_n^{[\mu}p_m^{\nu]}\,,
\end{equation}
where we have taken into account the fact that only the part of $A$ that is antisymmetric in $n$ and $m$ survives in the sum.
Eq.~\eqref{Jradscalar} is the equation for the angular momentum carried away by ``radiated'' massless scalars with exactly zero-frequency, in particular it can be specialized to dilaton emissions by taking $g_n = -\kappa\,m_n^2/\sqrt2$.

\subsection{Graviton}

Let us now turn to the case of the graviton field. We can write the angular momentum operator in De Donder gauge as follows
\begin{equation}\label{Jalphabetagrav}
	J_{\alpha\beta} = -i\int_{\vec k} a_{\mu\nu}^\dagger(k) 
	\left(
	P^{\mu\nu,\rho\sigma} k^{\phantom{[}}_{[\alpha} \frac{\overset{\leftrightarrow}{\partial}}{\partial k^{\beta]}}
	+
	2 \eta^{\mu\rho}\delta_{[\alpha}^\nu \delta^{\sigma}_{\beta]}
	\right)
	a_{\rho\sigma}(k)\,,
\end{equation}
where 
\begin{equation}\label{}
	P^{\mu\nu,\rho\sigma} = \frac{1}{2}\left(
	\eta^{\mu\rho}\eta^{\nu\sigma} + \eta^{\mu\sigma}\eta^{\nu\rho} - \eta^{\mu\nu}\eta^{\rho\sigma}
	\right)
\end{equation}
is the tensor structure appearing in the gauge-fixed De Donder action.
Defining 
\begin{equation}\label{}
 	\mathcal J_{\alpha\beta}= \langle  J_{\alpha\beta}\rangle\,,
\end{equation}
we thus find 
\begin{equation}\label{JabTT}
	\mathcal J_{\alpha\beta}
	=
	-i
	\int_{\vec k}
	f^\ast_{\mu\nu} \left(
	\eta^{\nu\rho}
	k^{\phantom{I}}_{[\alpha}\frac{\overset{\leftrightarrow}{\partial}}{\partial k^{\beta]}}
	+2\delta^{\nu}_{[\alpha}\delta^{\rho}_{\beta]}
	\right)
	f^{\mu}_\rho\,, 
\end{equation}	
where $f^{\mu\nu}$ involves transverse-traceless projectors given by sums over physical polarisations, i.e.
\begin{equation}\label{}
	f^{\mu\nu} = \Pi^{\mu\nu}_{\rho\sigma} F^{\rho\sigma}\,,\qquad
	F^{\mu\nu} = \sum_n \frac{\kappa p_n^\mu p_n^\nu}{p_n\cdot k-i0}\,,
\end{equation}
with
\begin{equation}\label{TTprojector}
	\Pi^{\mu\nu,\rho\sigma} = \sum_{j} \varepsilon^{\mu\nu}_j \varepsilon^{\ast \rho\sigma}_j =\frac{1}{2}\left(
	\Pi^{\mu\rho}\Pi^{\nu\sigma} + \Pi^{\mu\sigma}\Pi^{\nu\rho}-\frac{2}{D-2}\,\Pi^{\mu\nu}\Pi^{\rho\sigma}
	\right),
\end{equation}
and
\begin{equation}\label{Pimunu}
	\Pi^{\mu\nu} = \eta^{\mu\nu}+\lambda^\mu k^\nu + \lambda^\nu k^\mu\,,
	\qquad
	\lambda^2=0\,,\qquad
	\lambda\cdot k = -1\,.
\end{equation}

At this stage, we would like to get rid of the dependence on the reference vector $\lambda^\mu$, but this is not straightforward due to the derivative with respect to $k$ in \eqref{JabTT} and to the specific tensor structure contracting the two factors, which involves some free indices.
Let us therefore discuss the orbital and the spin angular momentum separately: $\mathcal J_{\alpha\beta} = \mathcal L_{\alpha\beta} + \mathcal S_{\alpha\beta}$,
\begin{equation}\label{LS}
		\mathcal L_{\alpha\beta}
	=
	-i
	\int_{\vec k}
	f^\ast_{\mu\nu}
	k^{\phantom{I}}_{[\alpha}\frac{\overset{\leftrightarrow}{\partial} f^{\mu\nu}}{\partial k^{\beta]}}
	\,,\qquad
	\mathcal S_{\alpha\beta}
	=
-	i
	\int_{\vec k}
	2f^\ast_{\mu[\alpha}
	f^{\mu}_{\beta]}\,.
\end{equation}
Relying only on the antisymmetric structure of the operator and on the exact transversality property\footnote{See \cite{Heissenberg:2024umh} for a generalization of this derivation in which $k_\mu F^{\mu\nu}=R^{\nu}$ is kept arbitrary.} $k_\mu F^{\mu\nu}=0$,
we find
\begin{equation}\label{}
	i\mathcal L_{\alpha\beta} =\int_{\vec k}
	\left(
	F^\ast_{\mu\nu} k^{\phantom{I}}_{[\alpha}\frac{\overset{\leftrightarrow}{\partial} F^{\mu\nu} }{\partial k^{\beta]}} 
	- \frac{F'^\ast}{D-2}
          k^{\phantom{I}}_{[\alpha}\frac{\overset{\leftrightarrow}{\partial}  F'}{\partial k^{\beta]}}
	+2 k^{\phantom{\ast}}_{[\alpha} F^\ast_{\beta]\mu}\lambda\cdot F^\mu 
	- 2 \lambda\cdot F^{\ast}_\mu k^{\phantom{\ast}}_{[\alpha} F_{\beta]}^\mu
	\right),
\end{equation}
where $F' = \eta_{\rho\sigma} F^{\rho\sigma}$, and
\begin{equation}\label{}
	i\mathcal S_{\alpha\beta} = \int_{\vec k}
	\left(
	2F^{\ast\phantom{\mu}}_{\mu[\alpha}F^\mu_{\beta]}  
	+  2k^{\phantom{\mu}}_{[\beta}F^{\ast\phantom{\mu}}_{\alpha]\mu}\lambda\cdot F^\mu
	+
	2\lambda\cdot F^\ast_{\mu}k^{\phantom{\mu}}_{[\alpha} F^\mu_{\beta]}
	\right).
\end{equation}
Therefore, the $\lambda$ dependence drops out in the sum and we obtain the remarkably simple expression
 (which also very recently appeared in \cite{Manohar:2022dea})
\begin{equation}\label{JIJgraviton}
	\mathcal J_{\alpha\beta} = -i\int_{\vec k}
	F^\ast_{\mu\nu} \left[
	\left(
	\eta^{\mu\rho}\eta^{\nu\sigma}-\tfrac1{D-2}\,\eta^{\mu\nu}\eta^{\rho\sigma}
	\right)
	k^{\phantom{I}}_{[\alpha}\frac{\overset{\leftrightarrow}{\partial}}{\partial k^{\beta]}}
	+2\eta^{\mu\rho}\delta^{\nu}_{[\alpha}\delta^{\sigma}_{\beta]}
	\right]
	F_{\rho\sigma}\,.
\end{equation}
Similarly, the two terms of \eqref{JIJgraviton} are not separately gauge invariant. Under $F_{\mu\nu}\to F_{\mu\nu}+\xi_{\mu} k_\nu + \xi_\nu k_\mu$, where now $k^\mu \xi_\mu=0$ in order not to spoil the transverse property $k^\alpha F_{\alpha\mu}=0$,
the trace parts are automatically invariant, $F' \to F' + 2\xi\cdot k=F'$, while
\begin{equation}\label{}
	\int_{\vec k}F^\ast_{\mu\nu} k^{\phantom{I}}_{[\alpha}\frac{\overset{\leftrightarrow}{\partial}}{\partial k^{\beta]}}
	F^{\mu\nu} 
	\to 
	\int_{\vec k} \left(F^\ast_{\mu\nu} k^{\phantom{I}}_{[\alpha}\frac{\overset{\leftrightarrow}{\partial}}{\partial k^{\beta]}}
	F^{\mu\nu}
	+ 2\xi\cdot F^\ast_{[\beta}  k^{\phantom{\ast}}_{\alpha]}	-k_{[\alpha} \xi^\ast \cdot F_{\beta]}
	\right)
\end{equation}
and 
\begin{equation}\label{}
	\int_{\vec k} 2 F^\ast_{\mu[\alpha} F^{^\mu}_{\beta]} \to \int_{\vec k} \left(2 F^\ast_{\mu[\alpha} F^{^\mu}_{\beta]}
	+2 \xi\cdot F^\ast_{[\alpha} k^{\phantom{\ast}}_{\beta]}
	+2k_{[\alpha}  \xi^\ast \cdot F_{\beta]}
	\right),
\end{equation}
so that the gauge-dependent terms only cancel out in the sum.
In view of these facts, while one can interpret the two quantities in \eqref{LS} as orbital and spin angular momentum, the same interpretation does not hold for the two terms in \eqref{JIJgraviton}.

Let us emphasize again that, although we focus here on the soft limit, Eq.~\eqref{JIJgraviton} only relies on the transversality of $F^{\mu\nu}$, and thus easily extends to the case where $F^{\mu\nu}$ is replaced by the more general $b$-space version of the $2\to3$ amplitude \cite{Jakobsen:2021smu,Mougiakakos:2021ckm,DiVecchia:2021bdo,Cristofoli:2021vyo,Cristofoli:2021jas}, a.k.a.~the stress-energy pseudotensor $\mathcal T^{\mu\nu}$ like in \cite{Manohar:2022dea}. Indeed, since an exactly transverse $\mathcal T^{\mu\nu}$ can always be achieved \cite{Kosmopoulos:2020pcd}, the steps leading from \eqref{LS} to \eqref{JIJgraviton} follow through.

The integrals that are needed to evaluate Eq.~\eqref{JIJgraviton} (for the soft case of interest here) are the same as for the scalar case. In particular, the new spin piece only gives an integral of the type $p_m\cdot K$ in \eqref{pnKint}.  We thus obtain our main result
\begin{equation}\label{Jgraviton}
	\mathcal J^{\alpha\beta} = \frac{G}{2} \sum_{n,m}
	\left[
	\left(
	\sigma_{nm}^2-\frac12
	\right)
	\frac{\sigma_{nm}\Delta_{nm}-1}{\sigma_{nm}^2-1} 
	-2\sigma_{nm}\Delta_{nm}
	\right](\eta_n-\eta_m)\,p_n^{[\alpha}p_m^{\beta]}\,.
\end{equation}
The symbols $\sigma_{nm}$ and $\Delta_{nm}$ are defined in \eqref{sigmasDeltas}.
We stress that the formula \eqref{Jgraviton} is Lorentz covariant and valid for arbitrary kinematics $p_n$ of the background hard process, i.e.~it hold regardless whether or not the outgoing momenta can be regarded as small deflections of the incoming ones. In fact, just like Weinberg's theorem, it holds independently of the number and of the specific details of the hard particles taking part in the background hard process: one need only assign their momenta. However, of course, it only captures the contribution to the angular momentum due to zero-frequency gravitons. We will explore its properties more in detail in Sect.~\ref{ssec:222}, discussing also its ``small $G\,$'' expansion valid in the PM regime.

\subsection{Vector and \texorpdfstring{$\mathcal N=8$}{N=8}}

Let us now briefly complement the scalar and graviton results discussed above with the analogous formulas for the massless vector, for which the soft factor takes the form
\begin{equation}\label{}
	f_j^\text{vec}(k)= \varepsilon^\ast_{\mu,j}(k) F^\mu(k)\,,
	\qquad
	F^\mu(k) = \sum_{n} \frac{\eta_n e_n \,p_n^\mu}{p_n\cdot k-i0}\,.
\end{equation}
Adopting Feynman gauge, the relevant operator is in this case
\begin{align}\label{Jvec}
	J_{\alpha\beta}^\text{vec} = -  i\int_{\vec k} a_{\mu}^\dagger(k) 
	\left(
	\eta^{\mu\nu} k^{\phantom{[}}_{[\alpha} \frac{\overset{\leftrightarrow}{\partial}}{\partial k^{\beta]}}
	+
	\delta_{[\alpha}^\mu \delta^{\nu\phantom{\mu}}_{\beta]}
	\right)
	a_{\nu}(k)\,,
\end{align} 
and translates into the classical average
\begin{align}\label{JIJvecsc}
	\mathcal J^\text{vec}_{\alpha\beta} = - i \int_{\vec k}
	F^\ast_{\mu} \left(
	\eta^{\mu\nu}
	k^{\phantom{I}}_{[\alpha}\frac{\overset{\leftrightarrow}{\partial}}{\partial k^{\beta]}}
	+\delta^{\mu}_{[\alpha}\delta^{\nu\phantom{\mu}}_{\beta]}
	\right)
	F_{\nu}\,.
\end{align} 
Like for the graviton, \eqref{JIJvecsc} results after a non-trivial cancellation of $\lambda$-dependent contributions between the orbital and spin angular momentum terms in \eqref{Jvec}. Similarly, the two terms in \eqref{JIJvecsc} are not separately gauge invariant. 
The integrals are the same as for the scalar and the graviton, and we obtain 
\begin{align}\label{Jradvector}
	(\mathcal J^\text{vec})^{\alpha\beta}
	=
	\frac1{16\pi}
	\sum_{n,m}
	\frac{e_n e_m}{m_n m_m}\left[ -
	\sigma_{nm}\, \frac{\sigma_{nm}\Delta_{nm}-1}{\sigma_{nm}^2-1}
	+ \Delta_{nm}
	\right]
	(\eta_n-\eta_m)\,p_n^{[\alpha}p_m^{\beta]}
	\,.
\end{align}

We are now in a position to discuss the total contribution to the angular momentum  in $\mathcal N=8$ supergravity from zero-frequency massless modes.
Using the couplings described below \eqref{fvecfsc}, we can combine dilaton and KK scalar contributions, which we can read off \eqref{Jradscalar}, and KK vector contributions dictated by \eqref{Jradvector} with the graviton one \eqref{Jgraviton}, obtaining the following simple result for the case of $\mathcal N=8$ supergravity:
\begin{equation}\label{JN8}
		\mathcal J^{\alpha\beta}_{\mathcal N=8}
		=
		\frac{G}{2}
		\sum_{n,m}
		\left[
		\tilde\sigma_{nm}^{\,2} \frac{\sigma_{nm}\Delta_{nm}-1}{\sigma_{nm}^2-1}
		-2 \tilde\sigma_{nm} \Delta_{nm}
		\right](\eta_n-\eta_m)p_n^{[\alpha}p_m^{\beta]}\,,
\end{equation}
where $\tilde\sigma_{nm}=\sigma_{nm}-1$ if $n$ and $m$ corresponds to states compactified along the same direction, so that $m_n = m_m$, while it equals $\tilde\sigma_{nm} = \sigma_{nm}$ otherwise.

\begin{figure}
	\centering
	\begin{tabular}{c c}
	\begin{tikzpicture}[scale=.78]
		\draw[help lines,->] (-4,0) -- (4,0) coordinate (yaxis);
		\draw[help lines,->] (.5,-3) -- (.5,3) coordinate (zaxis);
		\draw[help lines,->] (1,1) -- (-.9,-2.8) coordinate (xaxis);
		\draw[ultra thick, blue, ->] (-3,0) -- (3,0);
		\draw[ultra thick, ->] (-3,0) -- (-3,-2); 
		\draw[ultra thick, ->] (-3,0) -- (-2,-1.73205); 
		\draw[ultra thick, ->] (3,0) -- (3,2); 
		\draw[ultra thick, ->] (3,0) -- (2,1.73205);
		\draw[ultra thick, red, ->] (3,2) -- (2,1.73205);
		\draw[ultra thick, green!60!black, ->] (.5,0) -- (-.1,-1.2); 
		\draw (3,.7) arc (90:120:.7);
		\draw (-3,-.7) arc (-90:-60:.7);
		\node at(4,0)[right]{$y$};
		\node at(.5,3)[right]{$z$};
		\node at(-.9,-2.8)[left]{$x$};
		\node at(3,1.8)[right]{$-\vec p_1$};
		\node at(2.08,1.4)[left]{$\vec p_4$};
		\node at(-3,-1.8)[left]{$-\vec p_2$};
		\node at(-2.08,-1.4)[right]{$\vec p_3$};
		\node at(2.4,2)[above]{$\vec Q$};
		\node at(3,-.2)[below]{Particle $1$};
		\node at(-3,.2)[above]{Particle $2$};
		\node at(-.4,.4)[left]{$\vec{b}$};
		\node at(-.1,-1.2)[left]{$\vec{\mathcal J}$};
		\node at(2.71,1.1){$\Theta_{\!s}$};
		\node at(-2.7,-1.1){$\Theta_{\!s}$};
	\end{tikzpicture}
&
	\begin{tikzpicture}[scale=.78]
		\draw[help lines,->] (-4,0) -- (4,0) coordinate (yaxis);
		\draw[help lines,->] (.5,-3) -- (.5,2) coordinate (zaxis);
		\draw[help lines,->] (1,1) -- (-.9,-2.8) coordinate (xaxis);
		\draw[ultra thick, blue, ->] (-3,0) -- (3,0);
		\draw[ultra thick, ->] (-3,0) -- (-3,-3); 
		\draw[ultra thick, ->] (-3,0) -- (-2,-2.5); 
		\draw[ultra thick, ->] (3,0) -- (2,-.5);
		\draw[ultra thick, red, ->] (-2,-2.5) -- (-3,-3);
		\draw[ultra thick, green!60!black, ->] (.5,0) -- (-.1,-1.2); 
		\draw (-3,-.7) arc (-90:-65:.7);
		\node at(4,0)[right]{$y'$};
		\node at(.5,2)[right]{$z'$};
		\node at(-.9,-2.8)[right]{$x'$};
		\node at(3,-2)[below]{$\vec p_1^{\,\prime}=0$};
		\node at(2.08,-.5)[left]{$\vec p_4^{\,\prime}$};
		\node at(-3,-2)[left]{$-\vec p_2^{\,\prime}$};
		\node at(-2.08,-1.6)[right]{$\vec p_3^{\,\prime}$};
		\node at(-2.4,-2.9)[below]{$\vec Q^{\,\prime}=\vec p_4^{\,\prime}$};
		\node at(3,.2)[above]{Particle $1$};
		\node at(-3,.2)[above]{Particle $2$};
		\node at(-.4,.4)[left]{$\vec{b}$};
		\node at(-.1,-1.2)[left]{$\vec{\mathcal J}^{\,\prime}$};
		\node at(-3,-.7)[left]{$\Theta_{\!s}^{\,\prime}$};
	\end{tikzpicture}
\end{tabular}
	\caption{Coordinate choices in the two reference frames discussed in the text: the centre-of-mass frame (left panel) and the frame where particle 1 is initially at rest (right panel).}
	\label{scatteringplane}
\end{figure}
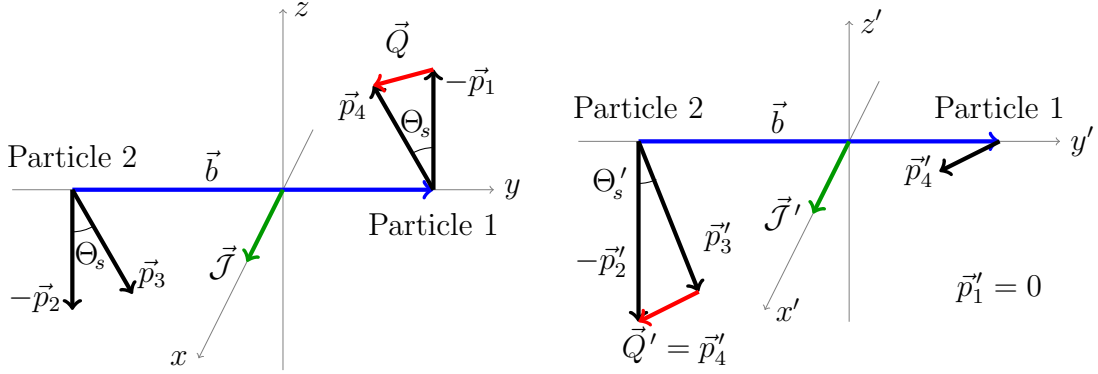

\subsection{\texorpdfstring{$2\to2$}{2->2} collisions}
\label{ssec:222}

Let us now specialize \eqref{Jgraviton} to the case of a $2\to2$ collision of spinless\footnote{Actually, Eq.~\eqref{Jgraviton} has been checked to capture the full $\mathcal O(G^2)$ angular momentum loss also for the more complicated case of hard particles with spin and generic spin alignments \cite{Alessio:2022kwv}.} point particles with masses $m_1$ and $m_2$, considering first the centre-of-mass frame (see Fig.~\ref{scatteringplane}), where 
\begin{equation}\label{CMkinematicsJ}
	\begin{aligned}
	p_1^\mu&=(-E_1,0,0,-p)\,,\qquad
	p_4^\mu=\left(E_1,0,-Q\cos\frac{\Theta_s}{2},+p-Q\sin\frac{\Theta_s}{2}\right),
	\\
	p_2^\mu&=(-E_2,0,0,+p)\,,
	\qquad
	p_3^\mu=\left(E_2,0,+Q\cos\frac{\Theta_s}{2},-p+Q\sin\frac{\Theta_s}{2}\right),
	\end{aligned}
\end{equation}
with $Q=2p \sin\frac{\Theta_s}{2}$ and so $Q\cos\frac{\Theta_s}{2}=p\sin\Theta_s$ and $p-Q\sin\frac{\Theta_s}{2}=p\cos\Theta_s$. To perform the calculation, let us recall the definition of the symbols $\sigma_{nm}$ and $\Delta_{nm}$ in terms of hard momenta in Eq.~\eqref{sigmasDeltas}. It is convenient to note that
$\sigma_{nm}=\sigma_{mn}$ and
\begin{equation}\label{eq:sigmamn4p}
	\sigma_{12} = \sigma_{34} = \sigma\,,
	\quad
	\sigma_{13} = \sigma_{24} =\sigma -\frac{Q^2}{2m_1m_2}\,,\quad
	\sigma_{14} = 1+\frac{Q^2}{2m_1^2}\,,\quad
	\sigma_{23} = 1+\frac{Q^2}{2m_2^2}\,.
\end{equation}
Clearly $\mathcal J^{xy}=\mathcal J^{xz}=0$, while we find
\begin{equation}\label{Jyz}
	\mathcal J^{yz}
	=
	G p\, Q \cos\frac{\Theta_s}{2}
	\sum_{n,m} \xi_{n,m}\left[ \left(
	\sigma_{nm}^2-\tfrac12
	\right) \frac{\sigma_{nm}\Delta_{nm}-1}{\sigma_{nm}^2-1}
	-
	2
	\sigma_{nm}\Delta_{nm}
	\right],
\end{equation}
where we introduced coefficients $\xi_{n,m} = \xi_{m,n}$ and the only non-zero ones are
\begin{equation}\label{}
	\xi_{1,4} = \xi_{2,3} = + 1\,,
	\qquad
	\xi_{1,3}=\xi_{2,4} = -1 \,.
\end{equation}

In the limit $Q^2\ll m_{i}^2$, i.e.~the standard PM regime, this result precisely agrees with \cite{Damour:2020tta,Manohar:2022dea} and is in fact proportional to $\operatorname{Im}2\delta$ (evaluated in the same regime),
\begin{equation}\label{lowenergymatching}
	\mathcal J^{yz}
	\sim
	\frac{4p}{Q}\,\lim_{\epsilon\to0}\left[
	-\pi\epsilon\operatorname{Im}2\delta
	\right] + {\cal O}(G^4),
\end{equation}
where $\operatorname{Im}2\delta$ is the imaginary part of the eikonal, whose divergent part reads
\begin{equation}\label{Im2delta}
	 \operatorname{Im}2\delta \simeq -\frac{G Q^2}{2\pi\epsilon} 
	\left[
	\frac{8-5\sigma^2}{3(\sigma^2-1)}
	+
	\frac{(2\sigma^2-3)\sigma\operatorname{arccosh}\sigma}{(\sigma^2-1)^{3/2}}
	\right].
\end{equation}
Note that, consistently with unitarity, the function within squared brackets in this equation is strictly positive for $\sigma>1$ so that $\mathcal J^{yz}$ in \eqref{lowenergymatching} corresponds to the system \emph{losing} angular momentum.
Moreover, since no energy-momentum is radiated to $\mathcal O(G^2)$, this angular momentum loss implies that the impact parameter becomes shorter, $\Delta b = - \mathcal J^{yz}/p$ to this order. 
In our approach, such an $\mathcal O(G^2)$ term emerges because we let the soft dressing \eqref{eq:eiksr} act on the full $S$-matrix, so that expectations like \eqref{SOS} can include interference terms between the $T$-matrix and the identity matrix.

The coefficient $4p/Q$ in \eqref{lowenergymatching} is precisely the one that ensures the agreement between two different formulas for the radiation-reaction correction to the deflection angle to 3PM order. One comes from an analyticity argument linking real and imaginary parts the 3PM eikonal $2\delta_2$ \cite{DiVecchia:2021ndb,DiVecchia:2021bdo}
\begin{equation}\label{}
	\operatorname{Re}2\delta_2^\text{RR} = \lim_{\epsilon\to0}\left[-\pi\epsilon\operatorname{Im}2\delta_2\right]
\end{equation}
so that
\begin{equation}\label{chiGPRC}
	\Theta_\text{3PM}^\text{RR} 
	= 
	-\frac{1}{p}\,\frac{\partial \operatorname{Re}2\delta_2^\text{RR}}{\partial b}
	= 
	\frac{2}{pb}\lim_{\epsilon\to0}\left[-\pi\epsilon\operatorname{Im}2\delta_2\right]
	\,,
\end{equation}
where we used the fact that the eikonal scales like $b^{-2}$ to this order.
The other one comes from the Bini-Damour linear response equation \cite{Bini:2012ji,Damour:2020tta,Bini:2021gat}, which to leading order reduces to
\begin{equation}\label{chiBD}
	\Theta_\text{3PM}^\text{RR} \simeq - \frac{1}{2p}\,\frac{\partial\Theta_{\text{1PM}}}{\partial b}\,\mathcal J^{yz}  \simeq \frac{Q}{2p^2b}\, \mathcal J^{yz} \,.
\end{equation}
So that \eqref{chiBD} agrees with \eqref{chiGPRC} thanks to \eqref{lowenergymatching}.

In the opposite limit $m^2_i\ll Q^2=s \sin^2\frac{\Theta_s}{2}$, the naive PM expansion breaks down \cite{Kovacs:1977uw,Kovacs:1978eu,DiVecchia:2022nna}. However, of course, Eq.~\eqref{Jyz} is still well defined and provides the smooth limit
\begin{equation}\label{}
	\mathcal J^{yz}
	\sim
	2Gs\,\sin\Theta_s\,\log\frac{\cos\frac{\Theta_s}{2}}{\sin\frac{\Theta_s}{2}}\,,
\end{equation}
which, when further expanded for small $\Theta_s$ yields
\begin{equation}\label{Jhigh}
	\mathcal J^{yz}
	\sim
	Gs\,\Theta_s\,\log\frac{4}{\Theta_s^2}\,.
\end{equation}
Comparing with the results of \cite{DiVecchia:2022nna} in the ultra-high-energy limit, and multiplying by the same factors which appear in \eqref{lowenergymatching}, we also get, to leading order in $G$,
\begin{equation}\label{Imhigh}
\frac{4p}{Q}\,\lim_{\epsilon\to0}\left[
-\pi\epsilon\operatorname{Im}2\delta
\right]
	\sim 
	G s \,\Theta_s\,\left[1+\log\frac{4}{\Theta_s^2}\right].
\end{equation} 
We see that the log-enhanced term in this equation is precisely the same as in \eqref{Jhigh}.
However, there is an additional non-log-enhanced term in \eqref{Imhigh}, which does not have a corresponding term in \eqref{Jhigh}.
We did not find similar relations between the two quantities in general, for intermediate values of the velocity.

We turn now to the mass dipole, $\mathcal J_{t I}$, using the kinematics \eqref{CMkinematicsJ} in the CM frame. While $\mathcal J^{tx}$ clearly vanishes, we get
\begin{equation}\label{}
	\mathcal J^{ty} = {G} Q \cos\frac{\Theta_s}{2}\sum_{n,m} c_{n,m} \left[
	\left(
	\sigma_{nm}^2-\tfrac12
	\right) \frac{\sigma_{nm}\Delta_{nm}-1}{\sigma_{nm}^2-1}
	-
	2
	\sigma_{nm}\Delta_{nm}
	\right] 
\end{equation}
where the only non-zero coefficients $c_{n,m} = c_{m,n}$ are
\begin{equation}\label{}
	c_{1,3} =  E_1\,,\qquad
	c_{1,4} = - E_1\,,\qquad
	c_{2,3} = E_2\,,\qquad
	c_{2,4} = -E_2\,.
\end{equation}
In the standard PM regime $Q^2\ll m_i$, one finds
\begin{equation}\label{signty}
	\frac{\mathcal J_{ty}}{b(E_1-E_2)}
	\sim
	\frac{\mathcal J_{yz}}{ 2 b p}\,.
\end{equation}
This result agrees with Eq.~(11) of \cite{Manohar:2022dea}, taking into account that the two frames are related by a rotation by $\pi/2$ about the $y$-axis and by the inversion $y\to-y$. On the other hand, the system's mass-dipole obeys the balance law $\Delta (b_1 E_1-b_2 E_2)=-\mathcal J_{ty}$, where $b_{1,2}$ are the impact parameters of each body with respect to the origin, so that $b=b_1+b_2$. Comparing with $\Delta(b_1+b_2)p=-\mathcal J_{yz}$ and substituting \eqref{signty}, leads to $\Delta b_1\, p= \Delta b_2 \, p= - \mathcal J_{yz}/2$. That is, each of the two bodies is responsible for ``half'' of the total angular momentum loss, regardless of their mass.

For the $tz$ component one gets instead
\begin{equation}\label{}
	\mathcal J^{tz} = {G} \sum_{n,m} d_{n,m} \left[
	\left(
	\sigma_{nm}^2-\tfrac12
	\right) \frac{\sigma_{nm}\Delta_{nm}-1}{\sigma_{nm}^2-1}
	-
	2
	\sigma_{nm}\Delta_{nm}
	\right] 
\end{equation}
where the only non-zero coefficients $d_{n,m} = d_{m,n}$ are
\begin{equation}\label{}
\begin{split}
	d_{1,3} &=  -Ep + E_1 Q \sin\frac{\Theta_s}{2} \,,\qquad
	d_{1,4} = -E_1 Q \sin\frac{\Theta_s}{2}
	\\
	d_{2,4} &= +E p - E_2 Q\sin\frac{\Theta_s}{2}\,, \qquad d_{2,3} = +E_2 Q \sin\frac{\Theta_s}{2}
\end{split}
\end{equation}
and thus, to leading PM order for $Q^2\ll m_i^2$,
\begin{equation}\label{signtz}
	\frac{\mathcal J_{tz}}{b(E_1-E_2)}
	\sim
	\frac{\Theta_s}{4}\,
	\frac{\mathcal J_{yz}}{b p}\,.
\end{equation}
In particular, $\mathcal J_{tz}\sim\mathcal O(G^3)$.
Like Ref.~\cite{Manohar:2022dea}, we do not find contributions to the mass dipole along the incoming direction to order $\mathcal O(G)$ and our approach only captures the very last term displayed in Eq.~(160) of \cite{Gralla:2021qaf}, while we do not obtain the remaining terms of Eqs.~(159), (160) of this reference. Such terms seem to come about due to the fact that the asymptotic behaviour of the particle trajectories is not the same as that of free particles, which is due to the long-range of the gravitational force in $D=4$. Comparison with \cite{Laddha:2018vbn,Saha:2019tub,Sahoo:2018lxl} indicates that such effects are captured by the $\log\omega$-corrected sub-leading soft theorem, which is compatible with us not finding them here: our approach is only based on Weinberg's leading soft theorem. Still, at least part of the mismatch could be related to a subtle difference between the standard centre-of-mass frame for the hard particles (which we adopted here) and the centre-of-energy frame for particles+field  \cite{Gralla:2021qaf,Gralla:2021eoi}.

Going to the frame where particle 1 is initially at rest (see Fig.~\ref{scatteringplane}) amounts to performing the following Lorentz transformation:
\begin{equation}\label{LorentzBoost}
	p'^\mu=\Lambda\indices{^\mu_\nu}\, p^\mu\,,\qquad
	\Lambda\indices{^\mu_\nu}
	= 
	\left(\begin{matrix}
		E_1/m_1 & 0&0 & -p/m_1\\
		0 & 1 & 0 & 0\\
		0 & 0 & 1 & 0\\
		-p/m_1 & 0 & 0 & E_1/m_1
	\end{matrix}\right).
\end{equation}
In particular, $\mathcal J'^{tz}=\mathcal J^{tz}$,
\begin{equation}\label{LtrJ}
	m_1\mathcal J'^{ty} = E_1 \mathcal J^{ty} + p \mathcal J^{yz}\,,\qquad
	m_1\mathcal J'^{yz} = p \mathcal J^{ty} + E_1 \mathcal J^{yz}\,,\qquad
	m_1p' = p E\,.
\end{equation}
Indeed, in this frame, $\mathcal J'^{yz}$ as calculated from \eqref{Jgraviton} is given by
\begin{equation}\label{Jprime}
	\mathcal J'^{\alpha\beta} =  \frac{G p E}{m_1}\,Q\cos\frac{\Theta_s}{2}  \sum_{n,m} \xi'^{n,m}
	\left[
	\left(
	\sigma_{nm}^2-\frac12
	\right)
	\frac{\sigma_{nm}\Delta_{nm}-1}{\sigma_{nm}^2-1} 
	-2\sigma_{nm}\Delta_{nm}
	\right],
\end{equation}
where now the only non-zero $\xi'^{n,m}=\xi'^{m,n}$ are
\begin{equation}\label{}
	\xi'^{2,3} =+ 1\,,\qquad \xi'^{2,4} =-1\,.
\end{equation}
Then we find, to leading order in the PM expansion,
\begin{equation}\label{}
	\frac{\mathcal J^{yz}}{p b} = 2 \, \frac{\mathcal J'^{yz}}{p' b}\,.
\end{equation}
This relation agrees with \cite{Manohar:2022dea}, and the above formulas are indeed consistent with the transformation law \eqref{LtrJ}, in particular with the overall sign in \eqref{signty}.

However, Eq.~\eqref{Jgraviton} goes beyond the leading PM order. For instance, expanding for $Q \simeq p (\Theta_{\text{1PM}}+\Theta_{\text{2PM}})$, the $\mathcal O(G^3)$ expansion of \eqref{Jprime} matches the $\mathcal D(\sigma)$ term in Eq.~(13) of \cite{Manohar:2022dea}, which arises in their calculation by interference terms between the $\delta(\omega)$ contribution and NLO deflection terms. 

\section{Connection with Other Approaches}

Let us now investigate the relation between the above expressions for $\mathcal J^{\mu\nu}$, in particular \eqref{JIJgraviton}, and more standard formulas that have been used for instance in \cite{Damour:2020tta,Jakobsen:2021smu,Mougiakakos:2021ckm}.
For the graviton case, we can start from Eq.~(2.2) of \cite{Damour:2020tta} cast in the form
\begin{equation}\label{JIJgravclass}
	\mathcal J_{IJ} = \int du \oint d\Omega(\hat x)
	\left[ 
	f_{\text{TT},\mu\nu}(u,\hat x)  \hat x_{[I} \partial_{J]} \partial_u f_\text{TT}^{\mu\nu}(u,\hat x)
	+ 
	2
	f_{\text{TT},\mu}^{[I}(u,\hat x)  \partial_u f_\text{TT}^{J]\mu}(u,\hat x)
	\right],
\end{equation}
where $f_{TT}(u,\hat x)$ is the position-space waveform normalized according to
\begin{equation}\label{softgravitonII}
	h^{\mu\nu} \sim \frac{1}{r}
	f_{\text{TT}}^{\mu\nu}
	\,.
\end{equation}
In the case of the memory waveform, comparing \eqref{hTT} with \eqref{softgravitonII},
\begin{equation}\label{memorymode}
	f_{\text{TT}}^{\mu\nu}(u,\hat x)
	= \Pi^{\mu\nu}_{\rho\sigma}(\hat x)
	\,\sum_{n}\frac{\kappa}{4\pi} \frac{p_n^\rho p_n^\sigma\, \theta(\eta_n u)}{E_n-\vec k_n\cdot\hat{x}}\,.
\end{equation}
On the other hand, focusing first on the orbital angular momentum, from \eqref{JabTT} we have
\begin{equation}\label{}
	i \mathcal L_{IJ} = \frac{1}{2}\int_{\vec k}
	\left[
	f^\ast_{\mu\nu} (k) k^{\phantom{I}}_{[I}\frac{\partial}{\partial k^{J]}}
	f^{\mu\nu}(k)
	-
	f^{\mu\nu} (k) k^{\phantom{I}}_{[I}\frac{\partial}{\partial k^{J]}}
	f^\ast_{\mu\nu}(k)
	\right],
\end{equation}
which we can cast in the following form using $f^\ast_{\mu\nu}(k)=-f(-k)$ and making the measure explicit,
\begin{equation}\label{}
	i\mathcal L_{IJ} = \frac{1}{2}\int\frac{d\Omega(\hat k)}{2(2\pi)^3}\int_{-\infty}^{+\infty} \omega\,d\omega\,
	f^\ast_{\mu\nu}(\omega,\omega\hat k)  \hat k^{\phantom{[]}}_{[I}\frac{\partial}{\partial \hat k^{J]}}
	f^{\mu\nu}(\omega,\omega\hat k)
	\,.
\end{equation}
Performing the inverse Fourier transform, and keeping track of the overall factor in \eqref{hmunufmunu},
\begin{equation}\label{}
	f^{\mu\nu}(\omega, \omega \hat x)
	=
	i4\pi
	\int_{-\infty}^{+\infty} du \,f^{\mu\nu}_\text{TT}(u,\hat x) \,e^{i\omega u}\,,
\end{equation}
one gets
\begin{equation}\label{}
	i\mathcal  L^{IJ} = \int d\Omega(\hat x)\int \,du'\,du\, f_{_{\text{TT}},\mu\nu}(u',\hat x)\, \hat  x^{\phantom{[]}}_{[I}\frac{\partial}{\partial \hat x^{J]}} (-i\partial_u)\int \frac{d\omega}{2\pi}\, e^{i\omega(u-u')}f_{\text{TT}}^{\mu\nu}(u, u\hat x)\,.
\end{equation}
Recognizing the derivative of the delta function, we find
\begin{equation}\label{}
	\mathcal L^{IJ} = \int d\Omega(\hat x)\int du\, f_{_{\text{TT}},\mu\nu}(u,\hat x)\, \hat  x^{\phantom{[]}}_{[I}\frac{\partial}{\partial \hat x^{J]}} \partial_uf_{\text{TT}}^{\mu\nu}(u, \hat x)\,,
\end{equation}
to see that this is exactly the orbital part of \eqref{JIJgravclass}. A similar manipulation goes through for the spin angular momentum as well, so that one sees that the $IJ$ component of \eqref{JabTT} and \eqref{JIJgravclass} are in fact the same. Still, subtleties related to the use of the standard formula \eqref{JIJgravclass} in frames where the centre of mass is not at rest have been pointed out \cite{Ashtekar:2017wgq,Ashtekar:2017ydh,Bonga:2018gzr,Manohar:2022dea}. This could explain the apparent clash between the covariance of $\mathcal J^{\mu\nu}$, the results obtained in the centre-of-mass frame \cite{Damour:2020tta,Gralla:2021qaf} and those found in the frame where particle 1 is initially at rest \cite{Jakobsen:2021smu,Mougiakakos:2021ckm}, as also discussed in \cite{Manohar:2022dea}. On the other hand the scope of applicability of \eqref{JIJgraviton} seems general, due to its manifest covariance \cite{Manohar:2022dea}.

As clearly emphasized in \cite{Veneziano:2022zwh}, when the canonical BMS gauge is chosen where the shear goes to zero as $u\to-\infty$ as in \eqref{Wmod}, namely
\begin{equation}\label{softgravitonIII}
	h^{\mu\nu} \sim \frac{1}{r}
	f_{\text{TT}}^{\mu\nu}\,,\qquad
	f_{\text{TT}}^{\mu\nu}(u,\hat x)
	=\frac{\kappa}{4\pi}\,\theta(u) \Pi^{\mu\nu}_{\rho\sigma}(\hat x)
	\,\sum_{n} \frac{ p_n^\rho p_n^\sigma\, \eta_n}{E_n-\vec k_n\cdot\hat{x}}
	\,,
\end{equation}
applying \eqref{JIJgravclass} ought to yield a vanishing result.
This  mechanism is indeed also manifest in the explicit calculation performed in Sect.~\ref{sec:angularmomentum} in $\omega$-space, as we now illustrate. In $\omega$-space, going to the canonical Bondi frame corresponds to changing the $-i0$ prescription as in \eqref{modi0} i.e. by sending $-i0\to-\eta_n i0$ in each term. Modifying accordingly \eqref{omegaintnm} and \eqref{omegaintmm}, we see that the first one vanishes identically, while the second one becomes $m$-independent and hence vanishes in the sum over $n,m$ by antisymmetry.
Thus, consistently with  the analysis in \cite{Veneziano:2022zwh}, we conclude that the contribution to $\mathcal J^{\mu\nu}$ that is captured by the present discussion can be in fact always set to zero by adjusting the BMS frame via \eqref{modi0}.

To conclude, let us briefly go back to another subtle issue, namely whether or not the $\mathcal J^{\mu\nu}$ discussed here should be regarded as radiated angular momentum. Our calculation and the one in \cite{Manohar:2022dea} seem to indicate that taking the expectation value of the graviton's angular momentum operator in the final state $\langle J_{\mu\nu}\rangle$ captures two types of contributions. One is the angular momentum carried to null infinity by propagating gravitons, whose frequency may be small but non zero, which starts at $\mathcal O(G^3)$. Another one is a contribution that is localized at the $\omega=0$ end of the spectrum, starts at $\mathcal O(G^2)$ in the PM expansion and is the only one captured by our formula \eqref{Jgraviton} (which is valid beyond the PM regime). As such, it may be more appropriate to interpret the latter as a static contribution that is stored in the gravitational field. Still let us emphasize that, if one's ultimate goal is to calculate the loss of mechanical angular momentum of the underlying two-body system via a balance law, only the sum of the two is relevant. In other words, the variation of mechanical angular momentum of the two body system is simply equal to $-\langle J_{\mu\nu}\rangle$, which automatically accounts for both radiative and static losses.

\subsection*{Acknowledgements} 
We would like to thank Francesco Alessio and Gabriele Veneziano for collaboration on closely related projects. RR would like to thank IHES for hospitality during the final part of this work and Thibault Damour for enlightening discussions.
The research of RR is partially supported by the UK Science and Technology Facilities Council (STFC) Consolidated Grant ST/T000686/1. RR would like to thank IHES for hospitality during the final part of this work. The research of CH (PDV) is fully (partially) supported by the Knut and Alice Wallenberg Foundation under grant KAW 2018.0116. Nordita is partially supported by Nordforsk.

\providecommand{\href}[2]{#2}\begingroup\raggedright\endgroup

\end{document}